\documentstyle[12pt,epsfig]{article}
\newcommand{\lsim}{\raisebox{-0.07cm}{$\, \stackrel{<}{{\scriptstyle
\sim}}\, $}}
\newcommand{\gsim}{\raisebox{-0.07cm}{$\, \stackrel{>}{{\scriptstyle
\sim}}\, $}}
\begin{document}
\thispagestyle{empty}

\begin{flushright}
DESY-97-233\\
November 1997
\end{flushright}
\vspace{2.5cm}

\begin{center}
{\LARGE\bf Physics with Polarized Protons at HERA}

\vspace{1cm}
{\large Albert De Roeck and Thomas Gehrmann}

\vspace*{1cm}
{\it Deutsches Elektronen-Synchrotron DESY, D-22603 Hamburg, Germany}
\vspace*{2cm}
\end{center}

\begin{center}
{\bf Abstract}
\end{center}

The operation of  HERA with polarized proton and electron beams  will 
allow to study a wide variety of observables in 
polarized electron--proton collisions at $\sqrt s=300$~GeV. 
The physics prospects of this project have been elaborated in detail 
in 
a dedicated working group, whose results we summarize in this 
report. We show that several important and 
often unique measurements in spin physics could be made at HERA. 
These include measurements of the polarized structure function 
$g_1(x,Q^2)$ at low $x$, a direct determination of the polarized 
gluon distribution $\Delta G(x,Q^2)$ for the region $0.002<x<0.2$ from 
polarized di-jet rates and hadrons with 
high $p_t$, polarized quark distributions 
from weak structure functions and semi-inclusive asymmetries, 
parton distributions in the 
polarized photon and information on the helicity
structure of possible new physics
 at large $Q^2$. HERA could therefore make a significant 
contribution to our understanding of spin effects in high energy collisions
and to the spin structure of the 
nucleon.

\vspace{5.5cm}
\noindent

\begin{center}
{\it Introduction to the Proceedings of the}\\ {\it  ``Workshop on 
Physics with Polarized Protons at HERA'', DESY, March--September 1997.}
\end{center}
\vfill

\setcounter{page}{0} 
\newpage

\begin{center}
{\LARGE\bf Physics with Polarized Protons at HERA}

\vspace{1cm}

{Albert De Roeck and Thomas Gehrmann}
\vspace*{1cm}

{\it Deutsches Elektronen-Synchrotron DESY, D-22603 Hamburg, Germany}
\vspace*{2cm}

\end{center}

\begin{abstract}
The operation of  HERA with polarized proton and electron beams  will 
allow to study a wide variety of observables in 
polarized electron--proton collisions at $\sqrt s=300$~GeV. 
The physics prospects of this project have been elaborated in detail 
in 
a dedicated working group, whose results we summarize in this 
report. We show that several important and 
often unique measurements in spin physics could be made at HERA. 
These include measurements of the polarized structure function 
$g_1(x,Q^2)$ at low $x$, a direct determination of the polarized 
gluon distribution $\Delta G(x,Q^2)$ for the region $0.002<x<0.2$ from 
polarized di-jet rates and hadrons with 
high $p_t$, polarized quark distributions 
from weak structure functions and semi-inclusive asymmetries, 
parton distributions in the 
polarized photon and information on the helicity
structure of possible new physics
 at large $Q^2$. HERA could therefore make a significant 
contribution to our understanding of spin effects in high energy collisions
and to the spin structure of the 
nucleon. 
\end{abstract} 

\section{Introduction}
\vspace{1mm}
\noindent
The commissioning of the HERA electron-proton collider 
(27.5~GeV electrons on 820~GeV protons)
five 
years ago opened up a completely new kinematical domain 
in deep inelastic scattering (DIS), and the two HERA experiments have provided 
a multitude of new insights into the structure of the 
proton and the photon since then. Some of the physics highlights 
from the first years of running are the measurement of the 
proton structure function $F_2(x,Q^2)$ at previously 
inaccessibly small values of $x$ (fraction of parton momentum inside 
the proton) and large values of $Q^2$ (negative
square of invariant momentum transfer from 
electron to proton), 
the production of jets in deep inelastic scattering and in photoproduction 
providing a new probe of the parton content of proton and photon, 
first studies of 
the proton structure at low $Q^2$ where a transition between 
hard and soft physics is expected, the observation and subsequent 
investigation of diffraction in deep inelastic scattering, 
and many more. It is therefore only natural to 
assume that the operation of HERA with polarized proton and electron 
beams~\cite{heraspin}
could add vital new information to our picture of the spin structure of the 
nucleon. 

The HERA electron beam is in fact naturally transversely polarized 
 due to the 
Sokolov-Ternov effect, and spin rotators can flip transverse into 
longitudinal polarization as needed for physics studies. The longitudinally
polarized electrons are already used in the HERMES~\cite{hermes} experiment, 
operating a fixed polarized nucleon gas target in the HERA electron 
beam ($\sqrt s=7$~GeV) to study polarized structure functions and 
semi-inclusive final states. The natural extension of this programme 
is to have a polarized proton beam, thus allowing to make a variety of 
physics studies in polarized electron-proton collisions at
a centre of mass system (CMS) energy of $\sqrt s=300$~GeV.

Polarization of the proton beam at HERA is technically more 
involved than for the electron beam, since protons do not 
polarize naturally in a storage ring. Polarized protons have to 
be generated with a high current polarized source 
(recent progress on polarized sources is reviewed in~\cite{hughes}), 
then 
accelerated from low energies while maintaining their polarization 
during the whole chain of pre-accelerators. Studies towards this 
challenging aim are ongoing, and so far the results look very encouraging. 
The technical aspects of this project are elaborated in~\cite{barber}. 
Based on these studies, it seems realistic to assume that 
HERA could be operated with polarized electron and proton beams, each  
polarized to about 70\%, reaching a luminosity of 
200 to 500~pb$^{-1}$  integrated over several years. 
Precision studies of polarized observables require moreover a 
good knowledge on the absolute polarization of proton and 
electron beams. Present day technologies, as used for example at HERMES,
allow to determine the polarization of the electron beam within about 4--5\%. 
Measuring the polarization of the high energy proton beam is on the other hand 
still a challenge. Various designs for  proton polarimeters have been 
proposed and are reviewed in~\cite{polari}, and it seems realistic that 
the proton polarization can be determined within about 5\% accuracy. 

Apart from protons, it would be also of interest to have data  on polarized
neutrons. Deuterium is  not a viable option for HERA, due to 
problems in rotating the transverse polarization into a longitudinal one
in the interaction regions, but $^{3}$He seems to be a good candidate.
For this workshop we have assumed a similar luminosity for polarized
$^{3}$He as for polarized protons.

The physics prospects of a polarized HERA collider 
were investigated for the first time in a working 
group of the 1996 ``Future Physics at HERA'' workshop~\cite{heraspin}.
The most important observables identified in this 
working group were the polarized structure function $g_1(x,Q^2)$, 
polarized weak structure functions, 
di-jet production in polarized DIS and polarized photoproduction of jets. 
The working group established 
the measurability of all these observables, given an integrated 
luminosity of at least 200~pb$^{-1}$.  

It was shown that a measurement of 
$g_1(x,Q^2)$ at HERA could decide between different predictions for 
the small-$x$ behaviour of this structure function, thus reducing the 
uncertainty on its first moment, the Ellis-Jaffe sum rule, from this 
region. Combining the HERA measurement at intermediate $x$
with present fixed target data at lower $Q^2$, it is moreover possible 
to put constraints on the polarized gluon distribution from 
the evolution of $g_1$. 
Another possibility to access $\Delta G(x,Q^2)$ is the study of the  
di-jet rate in polarized DIS. 
A study of spin asymmetries in charged current events 
would allow unique access to the polarized weak structure functions which 
probe different combinations of polarized quark distributions than 
their electromagnetic counterparts.
Finally,
polarized photoproduction of jets was found to be sensitive to 
the polarized gluon distribution of the proton in 
the forward  (proton) direction while 
testing the parton polarization in the photon in the backward 
(electron) direction.   

In addition to the physics programme at the polarized $ep$ collider,
it would be possible to study polarized 
proton-nucleon collisions in a fixed target experiment 
in the polarized HERA proton beam. This proposed 
experiment, presently called HERA--$\vec{{\rm N}}$, would 
require a polarized internal nucleon target and a 
dedicated new spectrometer. It could 
add numerous hadron-hadron observables~\cite{nowak}
to the HERA spin physics programme.
Very interesting results are expected here, which are fully
complementary to the RHIC~\cite{rhic} spin physics program. The two main issues
are the careful investigation of twist-3 effects in the transition
region between non-perturbative 
and perturbative QCD via single spin asymmetries and
the measurement of $\Delta G/G$ in photon and charmonium production via
double spin asymmetries.

The outcome of the previous workshop is however only the first 
step towards a spin physics programme at the HERA collider, since 
only a few promising channels had been identified and briefly investigated.
These channels have now to be further explored, 
including in particular 
realistic estimates of detector effects on potential observables. 
Moreover, one should expect that the physics programme 
of a polarized HERA collider would include many other interesting 
channels, which are yet to be identified.

Compared to the last workshop, progress has been made in various 
directions. It has been demonstrated that the asymmetries of 
jet rates in DIS and photoproduction as well as of polarized weak 
structure functions are only mildly affected by detector effects.
Moreover, the direct extraction of 
$\Delta G(x,Q^2)$ from simulated di-jet rates in polarized DIS 
has been demonstrated. This measurement of  $\Delta G(x,Q^2)$  can 
be accompanied by information from charged tracks at high $p_t$. Even 
tighter constraints on the polarized gluon distribution are obtained 
if data on $g_1(x,Q^2)$ and on di-jets are combined in a global fit. 
It is moreover illustrated that HERA can probe 
$\Delta G(x,Q^2)$ at $x$ values at least
one order of magnitude below any 
other experiment, thus being less sensitive on extrapolations for 
a determination of the first moment. 

The possibility of studying spin asymmetries on transversely 
polarized protons has been briefly investigated on the example of 
the structure function $g_2(x,Q^2)$. The resulting asymmetries 
turned out to be at least one order of magnitude too small to 
be measured. It is expected that the same negative result 
holds for other transverse observables. 

Up to now, semi-inclusive asymmetries 
had only been considered as probe of the polarized quark distributions 
at fixed target experiments. It has now been demonstrated that their 
study at the HERA collider will be worthwhile and will yield additional 
information on the flavour decomposition of the nucleon spin.

An exhaustive study of observables in polarized photoproduction 
has shown that jet and inclusive hadron production will be the most 
promising channels in a study of $\Delta G(x,Q^2)$ and of the 
partonic content of the polarized photon, while 
several  other channels have been demonstrated to be 
unmeasurable. The total cross section 
for polarized photoproduction is moreover of interest, since it is 
related to the Drell-Hearn-Gerasimov sum rule.

The spin transfer in fragmentation processes, in particular
to self-analyzing $\Lambda$ particles is one of the new reactions
studied in the present workshop. It could be shown that new information on
fragmentation functions of quarks to $\Lambda$ baryons could be
obtained in photoproduction at HERA, already with unpolarized protons.
The study of target fragmentation at a polarized HERA would
 moreover be of vital importance, since
fundamental insight into the origin of the breaking of the
Ellis-Jaffe  sum rule
could be gained from leading hadron production in
the target fragmentation region.

A variety of diffractive  reactions 
are presently studied at the  unpolarized HERA collider. 
The present workshop has addressed for the 
first time the question of 
a possible spin dependence of diffractive cross sections,  
studying diffractive 
vector meson production and diffractive deep inelastic scattering. 
While the former 
reaction was found to yield only unmeasurably small asymmetries, 
there is good hope that the latter could provide a very clean probe
of perturbative approaches to diffractive reactions.

Finally, the anomalous excess of events at large $Q^2$ recently 
observed at HERA has triggered much interest in this kinematical region 
and motivated several interpretations invoking physics beyond the 
Standard Model as an explanation. Taking a new contact term 
at high $Q^2$ as an example, it has been demonstrated during the 
workshop which new information on the helicity structure of 
possible new interactions could be gained from 
studying spin asymmetries 
at large $Q^2$.

\section{The polarized structure function $g_1(x,Q^2)$}
\begin{figure}[t!]
\begin{center}
\epsfig{file=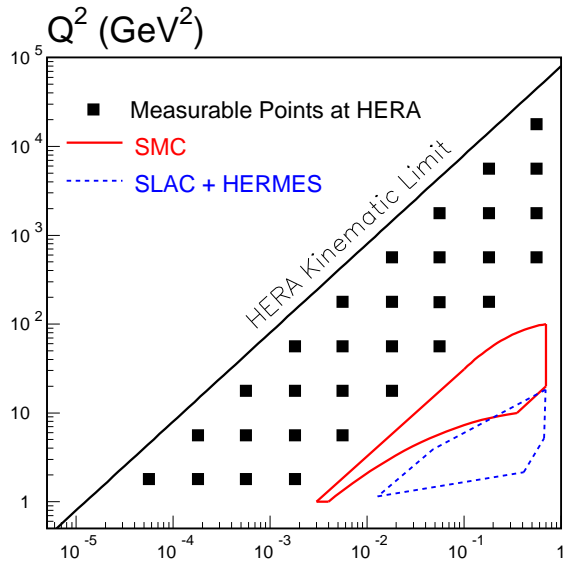,bbllx=0pt,bblly=40pt,bburx=500pt,bbury=390pt,width=10cm}
\caption{Measurable $x-Q^2$ region for a polarized HERA with the 
presently explored regions by CERN (SMC), SLAC and DESY (HERMES) 
experiments.}
\label{fig:kin}
\end{center}
\end{figure}

\vspace{1mm}
\noindent
The outstanding advantage of HERA is that it can measure 
structure functions at very small $x$ 
and very large $Q^2$. The kinematical reach is shown in
Fig.~\ref{fig:kin}, with a possible binning for measurements
of the polarized structure function $g_1(x,Q^2)$. In the quark parton model
$g_1$ can be interpreted as the (charge square weighted) density
of quarks with helicity parallel to the nucleon spin minus 
quarks with anti-parallel helicity.
The region covered by present fixed target experiments is
shown as well. HERA will extend the present region by two orders of magnitude 
both in $x$ and $Q^2$, reaching values of $Q^2$ up to 
$2\cdot 10^4$ GeV$^2$, and
values of $x$ down to $ 6\cdot 10^{-5}$. 
This highlights immediately  two very important 
contributions which HERA data can provide to the
understanding of the proton spin: the knowledge on the low-$x$ behaviour 
of $g_1$ and
the large  available $x-Q^2$ range, when including all polarized
experiments so far, which will allow for detailed QCD tests similar to ones 
in the unpolarized case.

\begin{figure}[t!]
\begin{center}
~ \epsfig{file=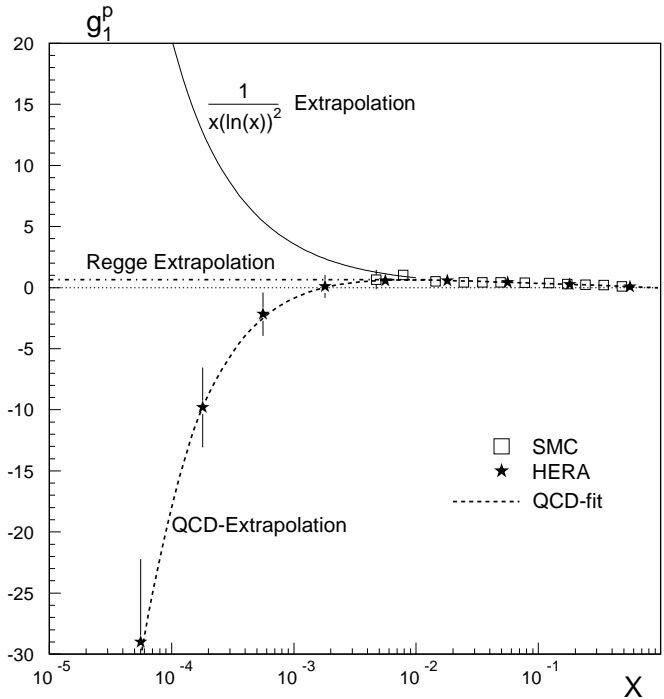,bbllx=0pt,bblly=0pt,bburx=560pt,bbury=560pt,width=10cm}
\caption{ The statistical uncertainty on the structure function $g_1$ of
the proton 
measurable at HERA, evolved to a value of 
 $Q^2 = 10$ GeV$^2$ for an integrated luminosity of
500 pb$^{-1}$, is shown. The SMC measurements are shown for comparison (for 
curves see text).}
\label{fig:g1p}
\end{center}
\end{figure}
\begin{figure}[t!]
\begin{center}
~ \epsfig{file=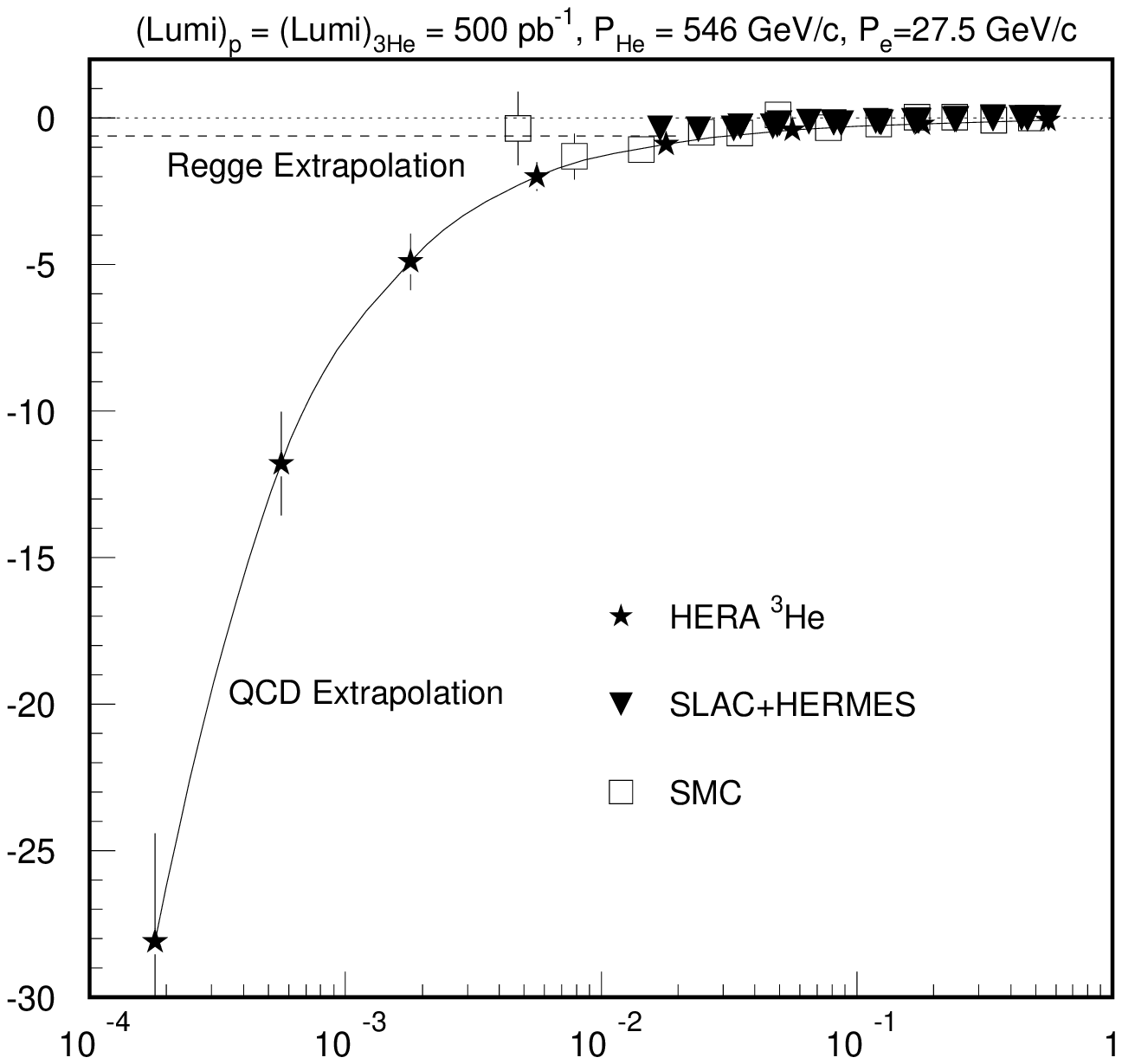,bbllx=0pt,bblly=200pt,bburx=600pt,bbury=650pt,width=14cm}
\caption{The statistical uncertainty on the structure function $g_1$ of
the neutron, using an $^3$He beam,
measurable at HERA evolved to  a value of
 $Q^2 = 10$ GeV$^2$ for an integrated luminosity of
500 pb$^{-1}$, is shown. 
Electron-neutron collisions are tagged via the remnant.
The fixed target measurements are shown for comparison (for 
curves see text).}
\label{fig:g1n_tagged}
\end{center}
\end{figure}
In particular the behaviour of structure functions at 
small $x$ has 
received much theoretical interest in recent years. While one 
observes on the one hand that ordinary perturbative evolution 
in $\ln Q^2$ yields a satisfactory description of the 
unpolarized proton structure down to values as low as 
$Q^2 = 1$~GeV$^2$, one would on the other hand expect terms 
proportional $\ln (1/x)$ to become important, thus rendering 
a different expansion and resummation parameter. 

The effects of small-$x$ resummation on the polarized structure 
function $g_1(x,Q^2)$ have been derived in~\cite{ber}, where it 
was shown that the most singular terms in the polarized structure 
function behave like $\alpha_s^n \ln^{2n}(1/x)$, while only terms   
proportional  $\alpha_s^n \ln^{n}(1/x)$ appear in the unpolarized 
case. First numerical studies, based on an insertion of the 
resummed splitting functions into the ordinary $\ln Q^2$ evolution 
equations were carried out in~\cite{blu} but the results  
so far are inconclusive.

Using a unified evolution equation, which incorporates both the 
conventional leading $\ln Q^2$ terms and the leading $\ln^{2}(1/x)$  
contributions, it was shown during the workshop~\cite{badelek} 
that resummation of the leading $\ln^{2}(1/x)$ terms  
should yield a sizable  
enhancement of the non-singlet content of the polarized structure 
function $g_1(x,Q^2)$ at small $x$. A 
prediction for the full structure function $g_1(x,Q^2)$, which contains 
both singlet and non-singlet components, is however not yet available in 
this unified approach. Given the behaviour of the $\ln^{2}(1/x)$ 
resummation at fixed $Q^2$~\cite{ber}, one would however expect 
even stronger enhancement effects in the singlet sector. 

Finally, the incorporation of infrared contributions from non-perturbative 
pion exchanges into the small-$x$ resummations~\cite{mana} allows to 
extend these spin-dependent
small-$x$ predictions to very low values of $Q^2$. 
This transition region between perturbative and non-perturbative 
physics has only recently become accessible at HERA. First measurements 
of the unpolarized proton structure at low $x$ and low $Q^2$ show 
that the infrared behaviour of the proton structure is governed by 
strong interaction dynamics as described in Regge theory.

Fig.~\ref{fig:g1p} shows the statistical precision of the measurement
of $g_1$ as a function of $x$~\cite{abhay}. 
Only the points with the highest 
$y$ values (lowest depolarization factor) are shown for each $x$ value.
The calculation is performed for 
an integrated luminosity of 500 pb$^{-1}$, $Q^2 > 1$
GeV$^2$, and the inelasticity range $0.01 < y < 0.9$.
The angle and energy of the scattered electron were required to be 
smaller than $177^0$ (defined with respect to the proton beam) and larger
than 5 GeV respectively. The radiative corrections have been 
studied in~\cite{bardeen} for the HERA kinematics, and are well 
under control.
Note that the expected asymmetries at HERA for $x \sim 10^{-4}$ are 
relatively small, about $10^{-3}$, which puts strong requirements
on the control of the systematic effects. Details 
on systematics are given in~\cite{abhay}.
The data points were centred on a curve which presents a low-$x$
QCD extrapolation resulting from a next to leading order (NLO) QCD fit
to the present fixed target data (see below). Other possible scenarios for the 
low-$x$ behaviour of $g_1$ are indicated in the figure:
the straight line is an 
extrapolation based on Regge phenomenology, and the upper curve
presents a scenario suggested in~\cite{abhay} where 
$g_1$ rises as $1/(x \ln^2(x))$, which is the maximally singular 
behaviour still consistent with integrability requirements.
All these scenarios are allowed by present day data from 
fixed target experiments.
Hence it is fair to say that we do not know the 
low-$x$ behaviour of $g_1$, and only HERA is able to solve this question,
like it did in the unpolarized case for the structure function $F_2(x,Q^2)$.

The problem of the unknown low-$x$ behaviour of the polarized 
structure functions is already very prominent in 
present polarized studies and questions, such as the measurement 
of the Bjorken sum rule~\cite{bj}. This is 
a fundamental sum rule due to isospin symmetry. It relates the 
 difference between the first moment of $g_1$ from 
proton and neutron 
to the weak coupling constants. 
Without QCD corrections it reads:
\begin{equation}
\Gamma_1^p - \Gamma_1^n =\int_0^1 g_1^p(x)~{\rm d}x - 
\int_0^1 g_1^n(x)~{\rm d}x
= \frac{1}{6} \left| \frac{g_A}{g_V}\right|
\end{equation}
where $g_A$ and $g_V$ are the axial and vector weak coupling constants of
neutron beta decay. QCD corrections up to $O(\alpha_s^3)$ have been computed
for this sum rule~\cite{qcdbj}.
This sum rule has been verified to about 10\% precision in present fixed 
target experiments, the largest uncertainty being due to the unknown 
behaviour of the polarized structure functions at low
$x$~\cite{smc,e155,altarelli}.
Hence only a significant improvement of the Bjorken sum rule measurement
can be expected when low-$x$ HERA data on $g_1$ become available.
The low-$x$  extrapolation  is at the same time the limiting factor 
for the determination of $\alpha_s(M_Z)$  
from the Bjorken sum rule~\cite{altarelli}, which consequently could be 
improved considerably as well with data from polarized HERA.

While the proton data at low $x$ by itself would be already very useful,
allowing to discriminate between somewhat extreme 
low-$x$ extrapolation scenarios as shown in  
Fig.~\ref{fig:g1p}, it would be very advantageous to have 
polarized low-$x$ neutron data as well. Those data  would additionally 
enable to measure
the singlet and non-singlet polarized structure functions at low $x$.
A study was made using polarized $^3$He at HERA~\cite{abhay}, from which 
$g_1^n$ can be extracted. The energy per nucleon for $^{3}$He is $Z/A$
times the proton energy, i.e.~546 GeV, reducing the kinematic reach
somewhat, but still allowing for measurements of $g_1$ down to $x= 10^{-4}$.
The dilution factor for $^3$He equals to 1/3. However, if the 
nucleus remnant can be
tagged downstream of the detectors, such that events can be selected which
correspond solely to electron-neutron
 scattering, this dilution factor can be bypassed.
Fig.~\ref{fig:g1n_tagged} shows the result for $g_1$ of the neutron,
measurable at HERA, for tagged events and an 
integrated  luminosity of 500 pb$^{-1}$. If it should turn out 
that remnant tagging
with high efficiency is not possible,
 the anticipated errors would increase 
by a factor $\sqrt 3$.  
Clearly this measurement is feasible and constitutes a strong encouragement
for the machine group to continue to study  this option.

The high quality  data from the fixed target experiments allows for 
quantitative QCD studies of the polarized structure function data, from 
which polarized parton distributions are extracted. 
In perturbative QCD structure functions are decomposed into convolutions
of perturbatively calculable coefficient functions and intrinsically 
non-perturbative
parton distributions which then vary with $Q^2$ according to perturbative 
evolution equations.
The present status of these studies is
reported in \cite{abhay}. Typically the singlet, non-singlet and gluon
distributions are extracted. The latter is of particular
interest. The violation of the Ellis-Jaffe sum rule 
in polarized deep inelastic 
scattering can be attributed  
to a large polarized gluon distribution 
and/or a negative polarized strange quark distribution,
depending on the chosen factorization scheme. 
Hence any information on the polarized gluon is vital for our full
understanding of the proton spin. Under evolution the singlet and 
gluon distributions mix, while the non-singlet evolves independently.
It turns out however that the QCD
fits constrain the gluon rather weakly, but some information on the first 
moment of $\Delta G$ can be obtained. The measurement from present 
day data gives $\int \Delta G(x)~{\rm d}x 
= 0.9 \pm 0.3({\rm exp})\pm 1.0({\rm theory})$ at $Q^2=1$~GeV$^2$.  
The theoretical error on this 
quantity is essentially dominated by the interpolation into 
the yet unmeasured low-$x$ region~\cite{altarelli}.
Including 
future HERA data will improve the experimental error to about 0.2.
The improvement in the theoretical error has not yet been quantified, but
it is expected that it will decrease by more than a 
factor 2 once $g_1(x,Q^2)$ is measured at low $x$.

In short the measurement of the polarized structure function $g_1$ at HERA
at low $x$ and large $Q^2$ is unique and 
 vital for future quantitative QCD studies of the spin structure of the
proton.

\section{The polarized gluon distribution $\Delta G(x,Q^2)$}
\vspace{1mm}
\noindent
From  the NLO
QCD fits discussed in the previous section, 
and from polarized parton density analyses in 
general, a  large polarized gluon distribution is suggested. 
The error on the polarized 
gluon from these fits is however large: 
although the first moment of the it can be determined with some accuracy,
there is still much freedom on its precise shape in $x$.
Moreover, it is crucial that these 
predictions are confirmed by direct experimental
test before the present standard interpretation of the data can be 
regarded as established.
Thus, 
important progress towards our understanding of the 
gluon contribution to the
spin structure 
of the proton can be made 
only
by direct measurements of $\Delta G$.
Polarized HERA is particularly suited for this task. It has been demonstrated
by the present unpolarized studies at HERA that the large CMS energy allows
for several processes to be studied which  show a clear sensitivity the gluon 
distribution in the proton. 
These processes include jet 
and high $p_t$ hadron production and  charm production 
both in DIS and photoproduction.
In this section we report on the studies on the extraction of  $\Delta G$
in deep inelastic scattering, 
while complementary studies on photoproduction data will be commented
upon in Sec.~5.

\begin{figure}[t!]
\begin{center}
\epsfig{file=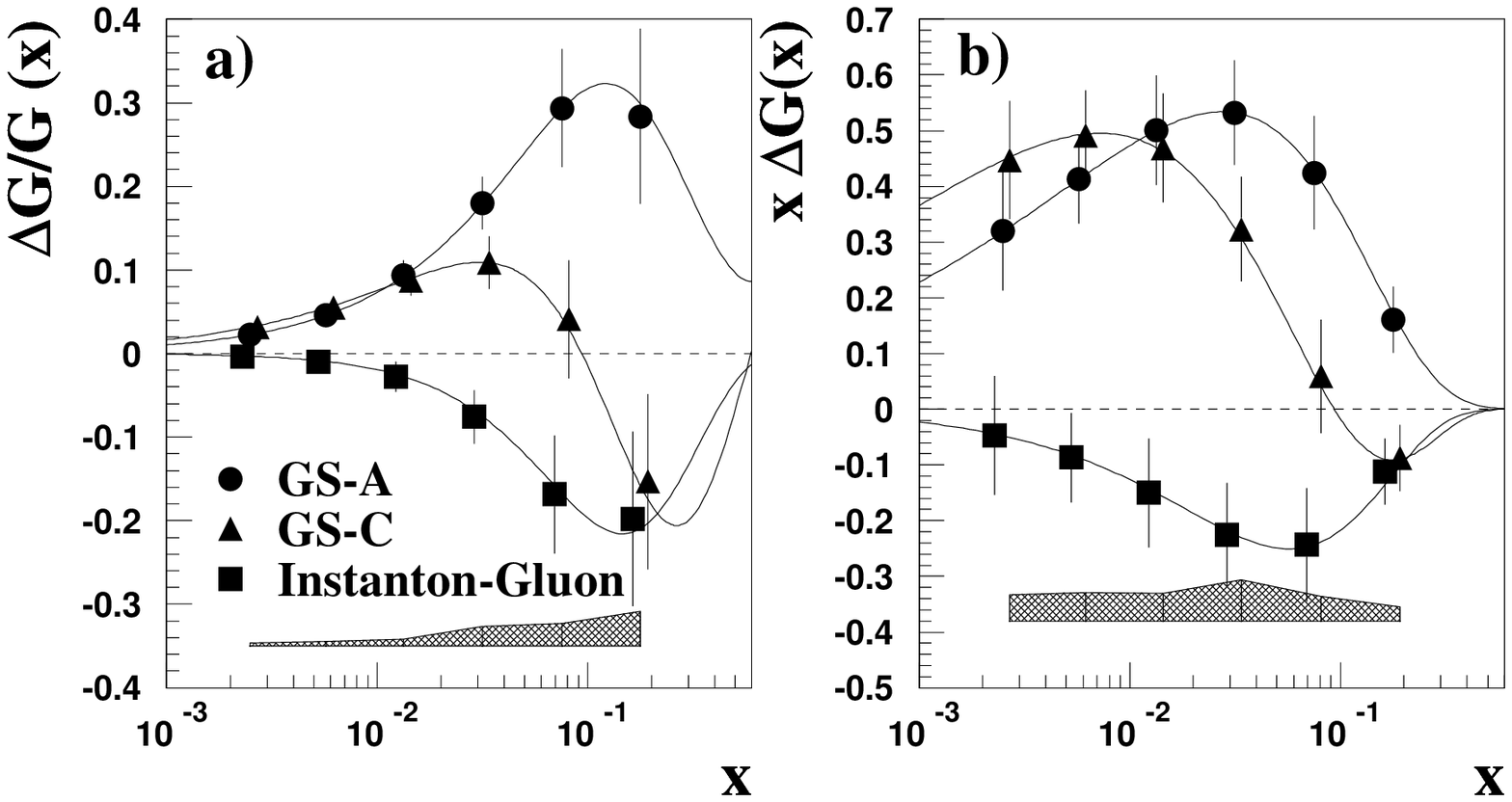,bbllx=0pt,bblly=220pt,bburx=650pt,bbury=600pt,width=15cm} 
\caption{Di-jets: Sensitivity to $\Delta G/G$ (a) and $x\Delta G$ (b)
 for  three different polarized gluon distributions shown
as solid lines and a luminosity
of 500~pb$^{-1}$, for $Q^2=20~{\rm GeV}^2$.}
\label{fig-g-500}
\end{center}
\begin{center}
\epsfig{file=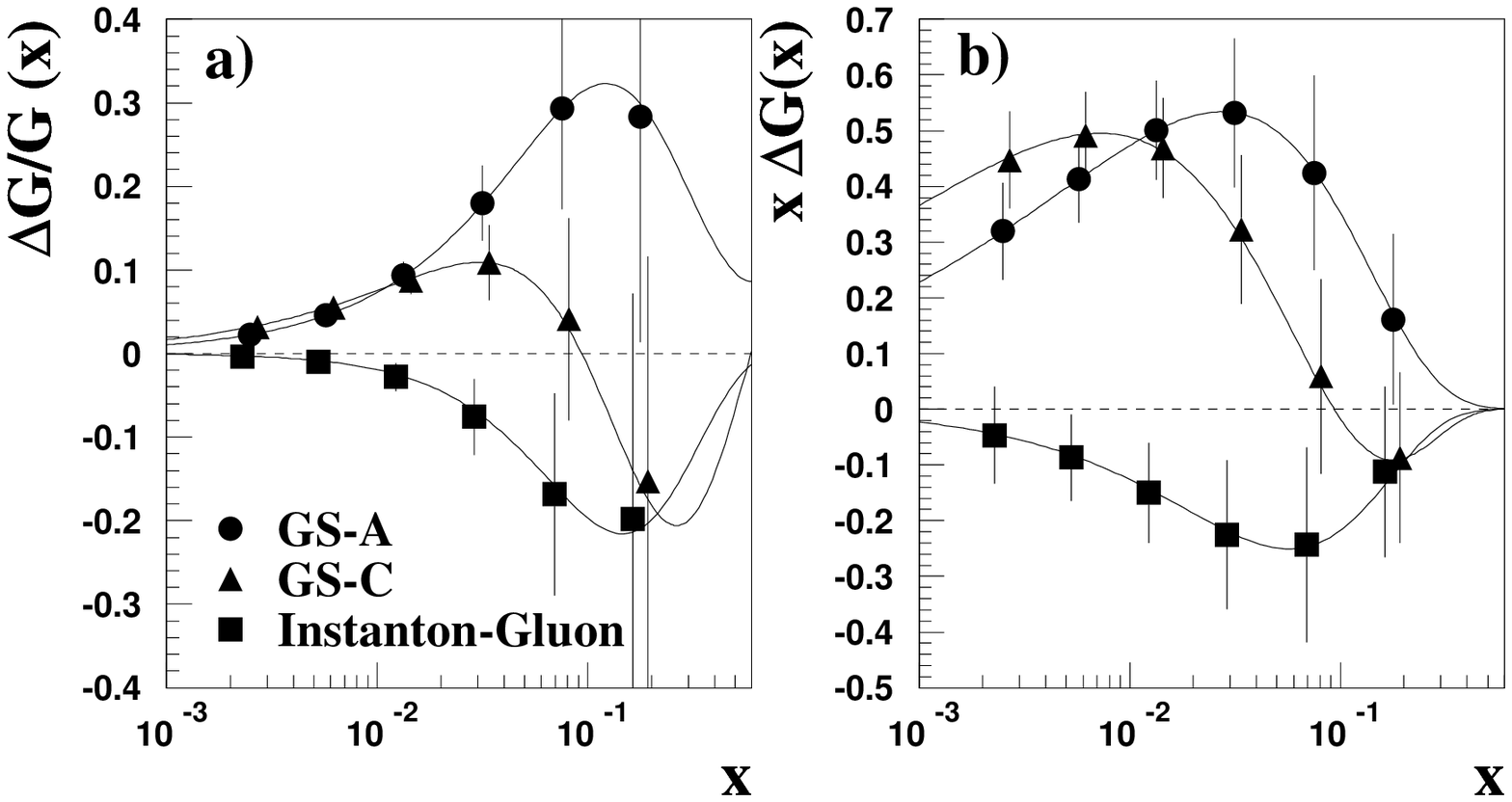,bbllx=0pt,bblly=220pt,bburx=650pt,bbury=600pt,width=15cm} 
\caption{High $p_t$ 
hadrons: Sensitivity to $\Delta G/G$ (a) and $x\Delta G$ (b)
 for  three different polarized gluon distributions shown
as solid lines and a luminosity
of 500~pb$^{-1}$, for $Q^2=20~{\rm GeV}^2$.}
\label{fig-t-500}
\end{center}
\end{figure}

The most promising process for  a direct 
extraction of  $\Delta G$ at HERA remains   di-jet production.
The underlying idea is to isolate boson-gluon fusion events, i.e.~a process
where the gluon distribution enters at the Born level. The exploratory 
leading order (LO) Monte Carlo 
study, reported in~\cite{raedel} which showed 
the  asymmetries at parton and
detector level, was further pursued~\cite{raedelnew}. 
The Monte Carlo generator {\tt PEPSI 6.5}~\cite{maul}
was extended with parton showers to approximate higher order effects. A more 
realistic detector simulation program~\cite{contreras} was used to check
the detector smearing effects on the asymmetries. Exact 
polarized NLO calculations 
were performed and compared with the LO ones~\cite{mirkes}: the NLO
QCD corrections were found to be moderate.
Finally a full unfolding of $\Delta G$ from the 
measured, background corrected (i.e.~QCD Compton events), asymmetries was made,
and the systematical errors were evaluated. The event sample used 
was selected with 
$5 < Q^2 < 100$~GeV$^2$ and $0.3 < y < 0.85$.  Jets are defined using
 the cone scheme, are
required to have a $p_t>5$~GeV and are restricted to the acceptance 
of a typical existing HERA detector by 
the requirement $|\eta^{jet}_{LAB}| < 2.8$, where
$\eta^{jet}_{LAB}$ is the pseudo-rapidity in the laboratory system.

The results are shown in Fig.~\ref{fig-g-500}. The measurable range in  
$x$ (of the gluon) is $0.002 < x < 0.2$. Statistical errors are shown for six
data points for three different assumptions on $\Delta G$, and the 
error band for the systematics is given. The assumed luminosity is 
500 pb$^{-1}$. The average $Q^2$ of this event sample 
 is very close to 20~GeV$^{2}$
therefore  results for  $\Delta G$ are presented at this value. 
The gluon  distributions are the Gehrmann-Stirling 
(GS) sets A and C~\cite{GS}, which
result from a QCD analysis of $g_1$ data, 
and a gluon distribution obtained from instanton calculations
\cite{kochelev}.
The distributions shown in Fig.~\ref{fig-g-500}, purposely selected,
indicate how poorly $\Delta G(x)$ is constrained by 
 the present polarized data. All of these distributions are
compatible with the  
available data, stressing the need for direct measurements
of $\Delta G(x)$. The $\Delta G(x)$ 
distribution extracted from the di-jet event 
is clearly able to judge between these scenarios.

We stress here that this measurement allows the determination of the 
{\it shape} of $\Delta G(x)$. Furthermore it reaches $x$ values lower than
any other measurement planned in future so far, and (for a GS-A type of gluon)
will measure about 75\% of the first moment $\int \Delta G(x)~{\rm d}x$.

Note that the gluon distribution will be also measured 
at RHIC~\cite{rhic} in polarized $pp$ collisions, in the range 
$0.03< x < 0.4$, with a comparable overall quality, but from 
an entirely different process (prompt photon + jet) with consequently
different systematic and theoretical errors. 
For RHIC the errors quoted~\cite{harrach}
on $\delta(\Delta G(x)/G(x))$ range from 
0.01 to 0.3, while for the di-jet measurement at HERA they range 
from 0.007 to 0.1.
In the range of overlap
HERA and RHIC are very complementary.
The fixed target experiment COMPASS~\cite{compass}
 at CERN is dedicated to the 
measurement of $\Delta G$. From the 
proposed charm measurement the expected precision~\cite{harrach} amounts to
$\delta(\Delta G(x)/G(x)) = 0.10$ for a (single) measurement covering 
the range $0.06<x<0.35$. At HERA, when taking the six data points 
together, we can achieve  $\delta(\Delta G(x)/G(x)) = 0.02$~\cite{raedelnew}
covering the range $0.0015<x<0.32$.

A different method to isolate photon-gluon fusion events at HERA
was investigated. Instead of tagging these events with two jets, two hadrons
with high transverse momentum $p_t$ opposite in azimuthal angle 
in the $\gamma^*p$ frame were required.
This method has recently been proposed for $\mu p$ polarized
fixed target experiments~\cite{bravar}. The {\tt PEPSI}
 Monte Carlo program was used,
and DIS events were selected in the same kinematic range as for the di-jets.
Two charged tracks with a $p_t$ larger than 1.5 GeV are required.
The resulting asymmetries at  hadron level are very similar to the ones
for the di-jet case~\cite{tracks}. 
The result of the unfolded gluon distribution is 
shown in Fig.~\ref{fig-t-500}. A similar level of discrimination power
as for the di-jet events is obtained, except in  the highest $x$ 
region, where the latter is superior.

We turn now back to the QCD fits on $g_1$ data for a moment.
 The poorly determined
gluon resulting from these fits suggests that one could gain substantially 
by combining the $g_1$ scaling violation information with the
direct measurement from the di-jets.
An exploratory study was made, using the values of $\Delta G(x)$
obtained from the di-jet 
analysis as an extra constraint in the fit.
The improvement of the errors on the first
moment of $\Delta G$  due to the inclusion of di-jet data
is shown in Table~\ref{tab-hera}.
\begin{table}[t]
\hfil
\begin{tabular}{||l|c||}
\hline\hline
  {\bf Analysis Type} &  {\boldmath $\delta (\int \Delta G~{\rm d}x)$} \\
\hline \hline
1. QCD  analysis of present $g_1$ data             &   0.3           \\
\hline
2. QCD  analysis of present \& projected  HERA $g_1$ data   &   0.2           \\
\hline
3. di-jets at HERA                   &   0.2          \\
\hline
4. combined 2 \& 3                    &    0.1   \\
\hline\hline
\end{tabular}
\hfil
\caption{\label{tab-hera} The expected statistical 
uncertainty in the determination of
  the first moment of the gluon distribution at $Q^{2} = 1$ GeV$^2$ 
using different information in a NLO QCD analysis.
 For the projected
data an integrated luminosity of 500~pb$^{-1}$ is assumed.}
\end{table}
 The first two  rows give the values quoted before, namely for the NLO QCD
analysis without and with  projected
HERA data for $g_1$. The third row shows the expected error if
only the di-jet asymmetry is added to the fixed target 
 $g_1$ data, and the fourth
row shows the total improvement using all available information.

In all, polarized HERA can make a very important contribution to the 
measurement of $\Delta G$ and hence to the understanding of the 
spin structure of the nucleon, in a unique
kinematic range. This constitutes therefore one of the major trump
cards for the physics case of polarized HERA.

\section{Polarized quark distributions}
\label{sec:pquark}
\vspace{1mm}
\noindent
The inclusive $g_1$ 
measurements from neutral current interactions ($\gamma^{\star}$ exchange),
discussed in Sec.~2, are sensitive to the sum of all quark flavours 
weighted by their charge squared. Neutron data in addition to proton 
data allow the extraction of singlet and non-singlet distributions.
To separate different quark flavours and valence quark distributions,
additional information is required.
\begin{figure}[t!]
\begin{center}
~ \epsfig{file=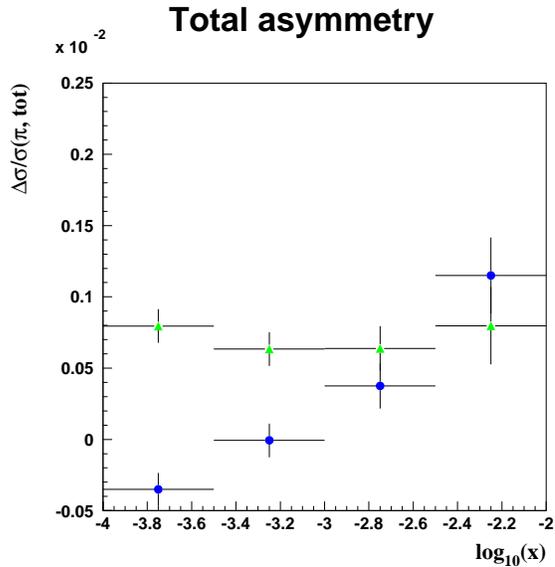,bbllx=0pt,bblly=0pt,bburx=560pt,bbury=650pt,width=8cm}
\caption{Total pion asymmetry $\frac{1}{P_eP_p}\frac{\Delta\sigma(\pi^+)
+ \Delta\sigma(\pi^-)}{\sigma(\pi^+)+ \sigma(\pi^-)}$ for 500 pb$^{-1}$
per relative polarization and $P_e = P_p = 0.7$, using {\tt PEPSI}. 
The triangles correspond to GS set A, the circles to GRSV-LO 
(STD)~\cite{GRSV} polarized parton densities.}
\label{fig:semi_incl}
\end{center}
\end{figure}
\begin{figure}[t!]
  \centering \epsfig{file=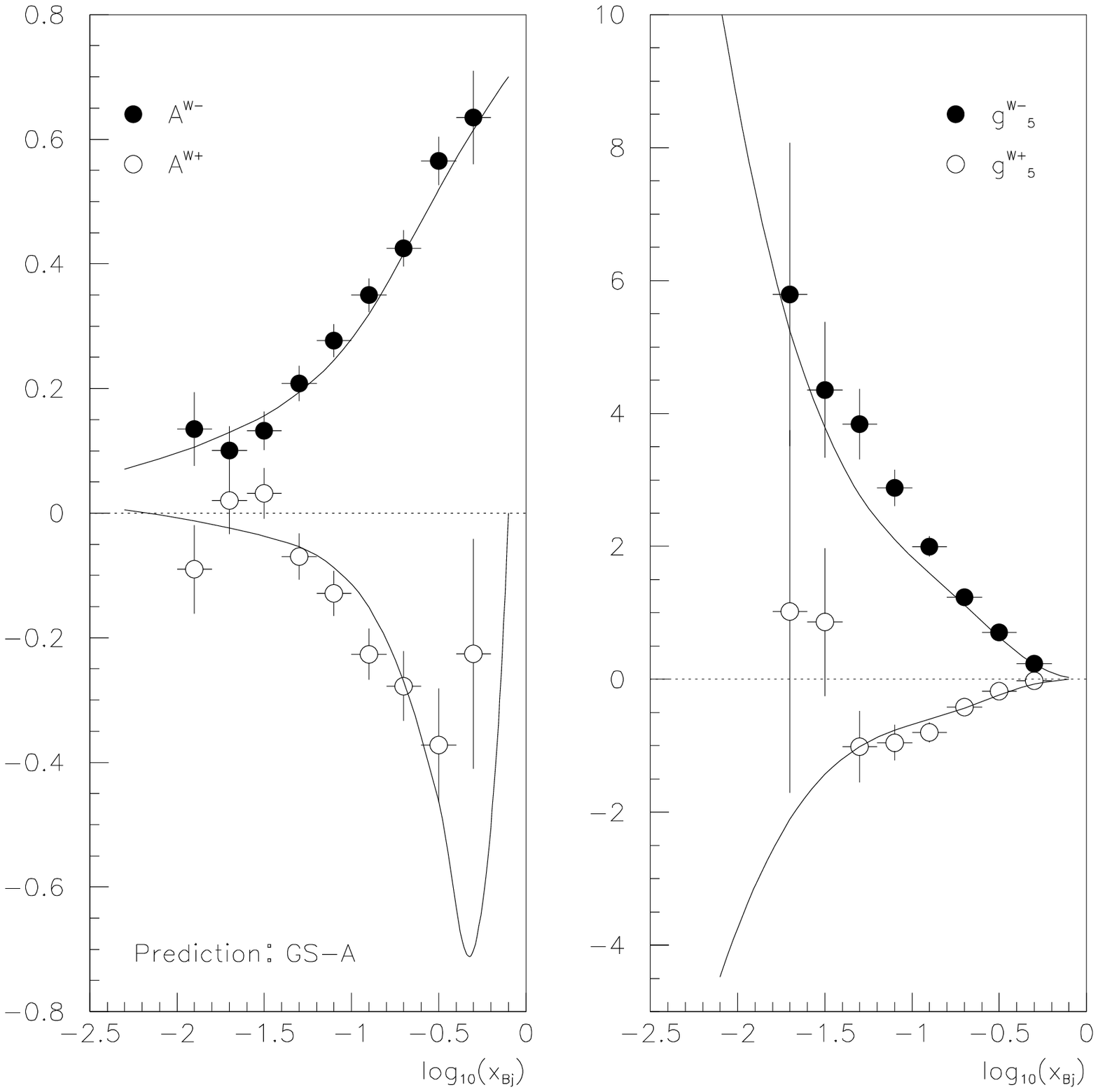,width=.75\textwidth}
\caption{Spin asymmetries $A^{W^-}$ (full circles, left side) and $A^{W^+}$
  (open circles, left side) for charged current 
events are presented for a total
  luminosity of 500 pb$^{-1}$. Also shown are the structure functions
  $g^{W^\pm}_5$ (right side) extracted from the asymmetries. The parton
  densities GS--A from \cite{GS} were used. The error bars represent
  the statistical uncertainty of the measurement.} 
\label{fig_results}
\end{figure}

Such information can be obtained from  semi-inclusive measurements, 
i.e.~measurements where a final state hadron is tagged. The aim is to 
select those particles which contain the quark which has been struck 
by the incoming boson. In practice one measures a convolution of the
fragmentation function with the parton densities, hence the study of
both topics is interconnected.

In~\cite{maul2} several issues on semi-inclusive 
measurements have been studied.
A {\tt PEPSI}
 Monte Carlo study was performed to check the purity of the so-called
favoured fragmentation functions (i.e.~when the fragmenting quark is 
flavour compatible with the hadron, like $u \rightarrow \pi^+$) as function of 
the classical fragmentation variable $z = P\cdot P_h/(P\cdot q)$. Here 
$P, P_h$ and $q$ are the four-vectors of the incoming proton, produced hadron
and exchanged boson, respectively. The purity is the probability that 
the hadron contains the struck quark or at least a quark with the 
same flavour as the struck quark. It is shown that the purity at HERA can 
reach 90\% at large $z$, hence the hadronic final state of HERA is a 
suitable environment for semi-inclusive studies.

Semi-inclusive asymmetries were studied for HERA at low $x$
($x<0.01$), by defining
asymmetries which use combinations of  $\pi^+$ and $\pi^-$ production.
By using either the sum or the difference of the $\pi^+$ and 
$\pi^-$ production asymmetries, it was shown~\cite{maul} that the 
 valence and sea quark contributions
can be disentangled at small $x$. A typical expected measurement of the 
semi-inclusive 
asymmetry is shown in Fig.~\ref{fig:semi_incl}, for a total integrated
luminosity of 1000 pb$^{-1}$.

While semi-inclusive pion measurements allow to separate the valence
and the sea contributions, one can distinguish positively and negatively
charged flavours via $W^{\pm}$ exchange, i.e.~via charged current interactions.
A study~\cite{maul2} shows that for an integrated luminosity of 
200 pb$^{-1}$
measurable asymmetries are obtained
for $W^-$ exchange, and that pion and kaon based asymmetries allow to
measure the relative importance of the spin contribution of $\bar{d}$
and $\bar{s}$ quarks, compared to that of the $u$ quark.

Another source of information is the inclusive measurement of charged 
current events. The asymmetry defined by 
\begin{equation}
A^{W\mp} =
\frac{d\sigma^{W^\mp}_{\uparrow\downarrow}-d\sigma^{W^\mp}_{\uparrow\uparrow}}
{d\sigma^{W^\mp}_{\uparrow\downarrow}+d\sigma^{W^\mp}_{\uparrow\uparrow}}
\nonumber\\
 = \frac{\pm 2bg^{W^\mp}_1+ag^{W^\mp}_5}{aF^{W^\mp}_1\pm
   bF^{W^\mp}_3}  
\approx \frac{g^{W^{\mp}}_5}{F^{W^{\mp}}_1} \label{eq_as}
\end{equation}
with $a = 2(y^2-2y+2)$
and $b = y(2-y)$, and $g^{W^-}_5 = \Delta u+\Delta c - \Delta\bar{d} - 
\Delta
\bar{s} $, $g^{W^+}_5 = \Delta d+\Delta s - \Delta\bar{u} - \Delta\bar{c}$.
A Monte Carlo study, including detector effects, was made for the 
measurements of the asymmetry and the extraction of $g_5$~\cite{contreras2}.
The total missing transverse momentum (which is a signal for the 
escaping neutrino) was required to be $P_{Tmiss}>15$ GeV,  and the region 
$Q^2>225$ GeV$^2$ has been selected for this analysis. This is a reasonable
assumption based on the present day experience at HERA.
The results for the asymmetries, including detector effects,
 are shown on the left side of Fig.~\ref{fig_results}. 
The error bars indicate the statistical precision 
of the measurement. The asymmetries are very large, as noticed before 
in~\cite{ansel}, so that the data allow for a significant 
measurement.
The solid line is the result of the exact analytical calculation of the 
asymmetry.  It shows that the detector smeared 
asymmetries  
are in good agreement with the true ones.
For the figure on the right side the approximation of  $A^{W\mp} =
g^{W^{\mp}}_5/F^{W^{\mp}}_1$
is tested.
The measured asymmetry has been multiplied with 
$F^{W^{\mp}}_1$
and   
 compared with   the analytical 
calculation for $g_5$. 
It shows that the approximation works well (to the 10-20\% level)
in our kinematic range. Hence these measurements can be used to 
extract e.g.~the $\Delta u$ and $\Delta d$ distributions at high $x$.

In addition to the above probes of different polarized 
 quark and anti-quark distributions, it might be possible to determine 
the total contribution (spin + angular momentum) of quarks 
to the proton spin from a study of Deeply Virtual Compton Scattering
(DVCS: $\gamma^* p \to \gamma p$). It has been recently suggested in the 
literature~\cite{dvcslit} that the DVCS cross section in 
unpolarized $ep$ collisions could be related to the total quark 
contribution to the proton spin. DVCS studies at low energies 
suffer from large QED background, which could be shown to be 
completely negligible at HERA~\cite{dvcs}. A measurement of this 
reaction appears therefore to be favourable already at the present 
unpolarized HERA collider.
 
\section{Photoproduction}
\vspace{1mm}
\noindent
Cross sections in electron-proton collisions become largest, if 
the virtuality of the  photon mediating the interaction is small. 
In this photoproduction limit, one can approximate the 
electron-proton cross section as a product of a photon flux factor and 
an interaction cross section of a real photon with the proton. 
Many unpolarized photoproduction reactions are presently 
measured at HERA, and their study has continuously improved our knowledge
on proton and photon structure as well as our understanding of the 
transition between real and virtual photons over the last years. 

A first investigation of polarized photoproduction 
has already been carried out during the last workshop~\cite{gammold}.
This study showed that photoproduction of single inclusive jets  is 
one of the most promising probes of both the polarized gluon distribution and  
the parton content of the polarized photon. Jet production in the 
 photon direction ($\eta_{LAB}\lsim 0$)
originates mainly from photon-gluon fusion
processes, and thus reflects the gluon polarization in the proton. The 
situation is more involved in the  proton direction 
($\eta_{LAB} \gsim 0$), where 
most events are induced by the yet unknown
resolved partonic content of the polarized photon. Given the polarized 
parton distributions in the proton to be known from other sources, jet 
photoproduction in the proton direction can be used to 
determine the polarized parton distributions in the resolved photon.  
An improved study of single inclusive jet photoproduction~\cite{gamma}
during the 
present workshop has shown that the sensitivity on the 
polarized photon structure is maximal for $\eta_{LAB} \gsim 2$, where 
still sizable jet rates guarantee small statistical errors on the 
expected asymmetries. 

A process very similar to single inclusive jet production is 
the single inclusive production of charged hadrons, which has been 
investigated in the present workshop for the first time~\cite{gamma}.
Although the 
production rate for individual hadrons is generally lower than the 
corresponding rate for jets, one expects a similar sensitivity 
on the polarized parton distributions in the photon and the proton from 
the measurement of this process, in particular since 
less stringent  kinematical 
cuts for single hadrons can be chosen.  
\begin{figure}[t!]
\begin{center}
~ \epsfig{file=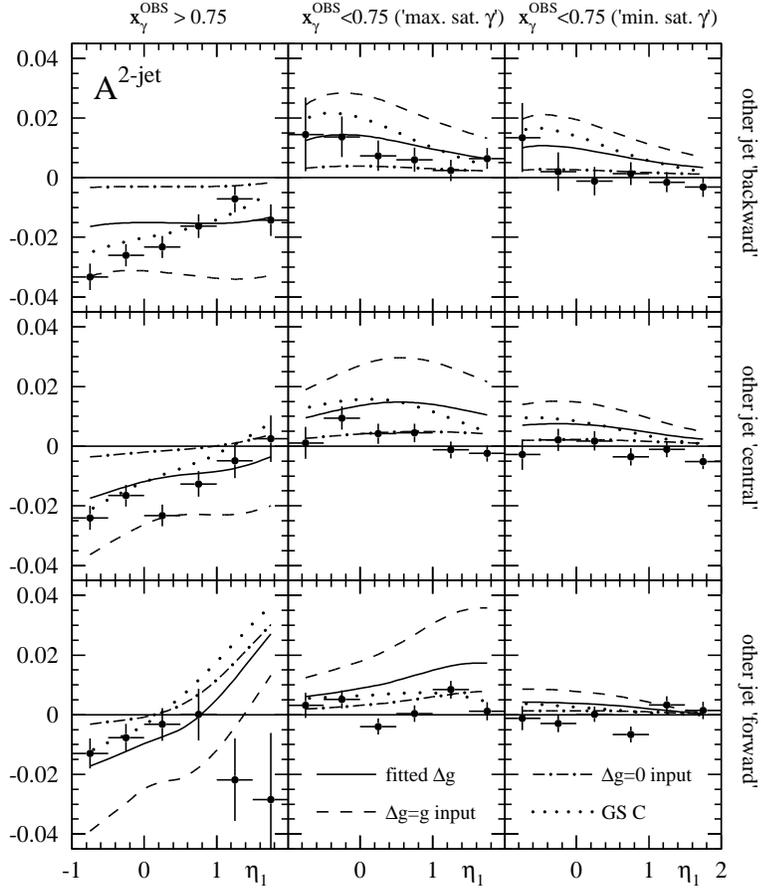,width=11cm}
\caption{Polarized photoproduction of di-jets ($E_{T,1} > 8$~GeV,
$E_{T,2} > 6$~GeV): 
asymmetries for direct (first column) and resolved (second and third 
column) photon contributions as function of the rapidity of the first 
jet and for different orientations of the second jet.
Second and third column correspond to different scenarios for 
the parton content of the polarized photon suggested 
in~\protect\cite{grvgamma}.
The error bars shown 
correspond to a Monte Carlo sample of 50~pb\protect{$^{-1}$}
for the GS(C) parton distributions and include parton showering,
hadronization, jet finding and jet clustering effects.}
\label{fig:phoprod}
\end{center}
\end{figure}

In the case of single inclusive jet and hadron production, it is not 
possible to make a clean separation between  direct and resolved photon 
contributions, the selection of a certain rapidity region only enhances 
or suppresses one of them. A better discrimination is possible, if
the inclusive production of di-jets is considered, since the di-jet 
rapidities allow (at lowest order) for a reconstruction of the 
incoming parton momenta, in particular to define 
the 'observed' parton momentum inside the photon $x_{\gamma}^{OBS}$. 
The resulting data can then be binned in $x_{\gamma}^{OBS}$, and one
usually attributes all events with $x_{\gamma}^{OBS}>0.75$ to  'direct' 
processes. The so-defined 'direct' sample has still a contribution 
from the resolved photon content, which is however small.

Polarized photoproduction of di-jets has been investigated in detail,
 including in particular effects due to 
parton showering, hadronization, jet finding and jet clustering. 
It could be demonstrated that, although 
these effects yield sizable corrections, the measurable asymmetry will
 largely be preserved at the hadron level~\cite{gamma}. 
An example for the correspondence of parton and hadron level asymmetries is 
shown in Fig.~\ref{fig:phoprod}, obtained with a moderate integrated 
luminosity of only 50~pb$^{-1}$. 
A direct determination of the polarized parton distributions in the 
photon from simulated data has up to now not been attempted. A 
first idea on the discriminative power of future measurements 
can however be gained by comparing the predictions obtained 
with the two (minimal and maximal) polarization scenarios 
proposed in~\cite{grvgamma}, as done in Fig.~\ref{fig:phoprod}.
Given the above
results for the di-jet case, it should be expected that the 
asymmetries in inclusive single jet production will survive 
at hadron level as well.

In summary, polarized photoproduction of jets and hadrons at HERA
has been proven to be a sensitive probe of polarized gluon 
distribution and parton content of the polarized photon.  
   Good results can already be achieved
with a rather low integrated  luminosity of 50~pb$^{-1}$. While the 
determination of the proton's polarized gluon distribution could be achieved 
from other processes at HERA or elsewhere as well, it must be emphasised 
that these processes are unique in probing the partonic content of the 
polarized photon.

In addition to the above processes, several other 
photoproduction channels, which are currently 
measured in unpolarized collisions at HERA, have been studied for 
the expected magnitude of polarization asymmetries. 
The most promising among these channels is the 
production of open charm, however the asymmetries 
will  become experimentally accessible only if 
the charm tagging efficiency can be improved considerably~\cite{gamma}. 
For the Drell-Yan process, large-$p_T$ 
photons \cite{gamma} and inelastic 
$J/\psi$ production~\cite{vecmes}, 
the situation is even worse since the production 
cross sections are 
relatively low, thus implying statistical errors larger than the anticipated 
asymmetries.

Finally, it should be pointed out  that a measurement of the total 
polarized photoproduction cross section $\Delta \sigma_{\gamma p} (\nu)$
as function of the photon-proton CMS energy $\nu$
at HERA would contribute to a precise determination of the 
Drell-Hearn-Gerasimov sum rule~\cite{dhg}. This fundamental sum rule, 
relating the total polarized photoproduction 
cross section to the anomalous magnetic moment of the nucleon, 
is presently tested in precision measurements at fixed target energies, 
which however rely on Regge-type extrapolations of $\Delta \sigma_{\gamma p} 
(\nu)$ for $\nu \to \infty$. A measurement of the polarized photoproduction 
cross section at HERA would test these Regge-theory predictions and  
put rigid constraints on the high energy contribution to the 
Drell-Hearn-Gerasimov sum rule.

\section{Spin effects in fragmentation}
\vspace{1mm}
\noindent
The polarized parton distributions that have been discussed in 
 detail above describe  
the probability of finding a parton of a particular species having its 
spin aligned or anti-aligned with the spin of the nucleon.
Correspondingly, one can define polarized fragmentation functions 
parameterizing the probability of a polarized parton fragmenting into a
hadron with spin aligned or anti-aligned to the parent parton spin. 
These polarized fragmentation functions are however experimentally 
 very hard to 
access for most hadrons, as they require the measurement 
of the spin state of a final state particle. Such a 
measurement is in 
practice only feasible for particles with dominant parity violating 
decay modes such as the $\Lambda$ baryon.

 First studies~\cite{fsv,bralambda} 
on the polarized 
fragmentation functions into $\Lambda$'s have been carried out 
recently. These studies consider three possible 
scenarios for the spin transfer to the $\Lambda$. A naive 
approach, based on the non-relativistic quark model
would predict that  
the $\Lambda$ spin is carried only by the $s$ quark, while the $u$ and 
$d$ quarks do not contribute to its spin. Secondly, in analogy to the 
Ellis-Jaffe sum rule for nucleon spin structure functions, one can 
deduce SU(3)$_f$-based relations between the polarized
$\Lambda$ fragmentation 
functions for different quark flavours~\cite{bi}. Assuming a breaking 
of this sum rule similar to the well known breaking of the Ellis-Jaffe 
sum rule, one obtains a positive contribution only from $s$ quarks, 
while $u$ and $d$ quarks contribute with negative sign to 
polarized $\Lambda$ fragmentation. Finally, it could  be possible that, 
in complete contradiction to the above models, the polarized 
fragmentation functions of $u,d,s$
quarks into $\Lambda$'s are simply identical. These three scenarios have 
been elaborated in~\cite{fsv}, where they have been imposed as 
low-scale boundary conditions for the perturbative evolution of the 
polarized fragmentation functions into $\Lambda$'s. 
\begin{figure}[t]
\begin{center}
~ \epsfig{file=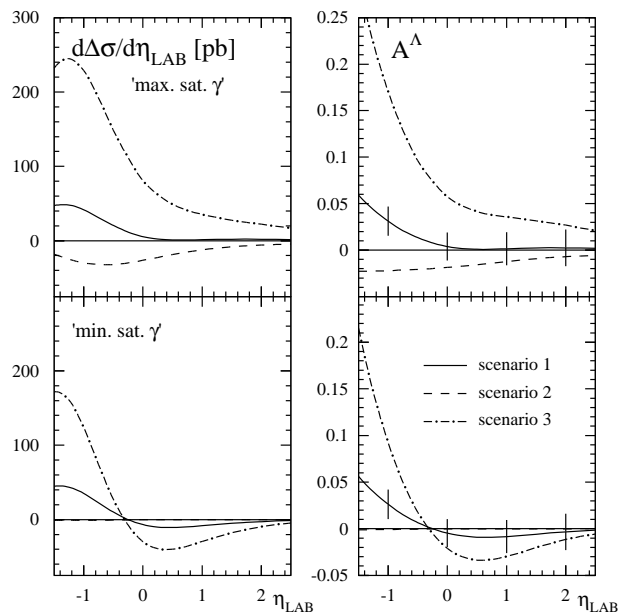,width=9cm}
\caption{Photoproduction of polarized $\Lambda$'s in collisions 
of polarized electrons on unpolarized protons: cross sections and 
asymmetries for $p_T>2$~GeV. Upper and lower row correspond to different 
parametrizations for the parton content of the 
polarized photon~\protect\cite{grvgamma}, the 
three scenarios for polarized $\Lambda$ fragmentation are explained in the 
text. Expected statistical errors are for an integrated luminosity of 
100~pb$^{-1}$ and a $\Lambda$ reconstruction efficiency of 0.1.}
\label{fig:lambda}
\end{center}
\end{figure}

It has been demonstrated~\cite{fsv} that 
a fit to LEP data only is insufficient to discriminate the 
above scenarios. Studying the production of polarized $\Lambda$'s in 
collisions of polarized electrons on unpolarized protons would 
on the other hand enable a distinction of the scenarios already with 
a moderate integrated luminosity of 100~pb$^{-1}$ 
and an assumed $\Lambda$ reconstruction efficiency of 0.1
in the 
photoproduction channel (see Fig.~\ref{fig:lambda}), more 
luminosity would allow a similar distinction from semi-inclusive $\Lambda$ 
production in DIS as well~\cite{fsv}.

It should be kept in mind that these studies of polarized $\Lambda$ 
fragmentation at HERA only require longitudinal polarization of
the electron beam, while the protons can be unpolarized. 
This programme could therefore
start already two years from now, once electron spin rotators 
are installed around the two HERA collider experiments.   
Polarization of the HERA proton beam would then give access to 
different combinations of asymmetries in polarized $\Lambda$ production. 
This case has up to now only been investigated~\cite{fsv}
for semi-inclusive 
deep inelastic scattering in view of a possible determination of the
strange quark polarization in the proton, which was found to be not feasible. 

Most studies of hadron fragmentation in deep inelastic scattering 
that have been carried out up to now were only concerned about the 
fragmentation of the current parton, that has been struck by the 
virtual photon. The fragmentation of the target remnant has on the 
other hand not received very much attention: it could  hardly be accessed 
in the fixed target experiments and  
until recently no theoretical models for 
it existed.
With the start of the HERA collider, where 
current fragmentation and target fragmentation take place in well 
separated regions of the detectors, the target fragmentation region has 
become accessible experimentally in principle. 
A possible theoretical description of phenomena in the target fragmentation 
region is given by the fracture functions introduced in~\cite{frac},
which parametrize the probability of tagging a particular hadron 
species in the target fragmentation region of a DIS event at fixed 
kinematics.

 A case of particular interest in spin physics is the 
configuration where a meson carrying almost the whole incident 
proton momentum is detected in the target fragmentation region
of polarized DIS. In this 
configuration, one would expect the fracture function to factorize into
a factor for the transition of a proton into a meson and another baryon 
(e.g.~$p\to \Delta^{++} \pi^-$) and the structure function of the 
baryon. The study of events with a highly energetic meson in polarized 
deep inelastic scattering would thus allow to access the spin structure 
functions of unstable baryonic excitations~\cite{fracph}. Such measurements
on various baryons would then enable the determination of 
the Ellis-Jaffe sum rule for a variety of baryon targets, thus testing whether 
the observed violation is indeed target independent and 
related to a fundamental property of the QCD vacuum, as suggested 
in~\cite{tindep}. This type of spin structure measurement 
from tagged mesons in the target fragmentation region would 
be unique at a polarized HERA collider. It requires however apart from
polarized protons also polarized neutrons ($^3$He), and a much 
improved instrumentation of the present detectors in the proton 
remnant direction.

To summarize, a variety of spin effects in fragmentation could be studied at 
HERA. The starting point of a spin physics programme could be marked 
by measurements of the polarized $\Lambda$ fragmentation functions, which 
requires only polarization of the electron beam.  
Once the proton beam is polarized also, several other observables 
become accessible.
Apart from measurements in the current fragmentation region of 
polarized deep inelastic scattering, which have been discussed 
already in Section~\ref{sec:pquark}, the target fragmentation 
region should receive particular attention: a measurement of 
forward meson production could make substantial contributions to 
 our understanding of the origin of Ellis-Jaffe sum rule violation.

\section{Diffractive processes}
\vspace{1mm}
\noindent
A sizable fraction of electron-proton collisions at HERA shows one remarkable 
feature in the hadronic final state: the incoming proton is 
either left intact or dissociates into a low mass state, 
separated by a large rapidity gap from the rest of the hadronic final state.
At HERA either the proton or a large rapidity gap can be observed.
These events are predominantly produced by a phenomenon termed
diffraction and occur for a variety of reactions. 
Examples are diffractive final states in deep inelastic scattering,
diffractive production of jets or heavy 
flavours and diffractive vector meson production. The first observation 
of diffractive phenomena shortly after the start of the HERA physics 
programme has triggered much theoretical effort towards an understanding 
of diffraction in electron-proton interactions. Although much progress 
both in the theoretical description and the experimental study 
of diffractive reactions has been made in the meantime, it is 
still fair to say that this phenomenon is not unambiguously understood 
at present, since it contains both perturbative and non-perturbative 
components. However, various 
theoretical descriptions  for diffractive reactions have been 
proposed, emphasising either the perturbative or non-perturbative 
components of the reaction cross section, a unified description is still 
due.

The spin dependence of diffractive reactions can in principle be 
studied both in perturbative and non-perturbative models. Studies in 
the present workshop have focused in particular on perturbative approaches,
in which predictions for the polarized cross sections 
in diffractive deep inelastic scattering and diffractive vector meson 
production were made. The non-perturbative contribution to spin-dependent 
diffractive reactions has been investigated~\cite{diff} in the 
framework of Regge theory. It could be demonstrated that diffraction at 
zero invariant momentum transfer off the proton receives only contributions 
from three particle cuts in Regge theory. The suppression factor between 
these cuts and the dominant unpolarized pomeron pole
exchange is identical 
to the suppression between polarized and unpolarized inclusive 
structure functions at small $x$. One expects however that the 
coupling factor of the three particle cut appearing in the 
 polarized cross section 
is small compared to the 
pole exchange appearing in the unpolarized cross section.  

Most perturbative approaches to diffractive reactions are based on 
the two gluon exchange model, which assumes that the 
diffractive reaction is mediated by the exchange of two 
gluons (in a color singlet state) between the proton and the 
virtual photon. The unpolarized diffractive cross section in this model 
is predicted to be proportional to the square of the unpolarized gluon 
structure function. An extension of the two gluon exchange model 
to spin-dependent diffraction 
has recently been discussed in~\cite{diff,ryskin}. In this approach, which 
takes up to now only the leading $\ln Q^2$ terms into account~\cite{ryskin}, 
one finds 
that the diffractive cross section is proportional to the product
of unpolarized and polarized gluon structure functions with 
a small admixture  from the polarized quark structure function.
For the fraction of diffractive events observed at 
small $x$, one should 
expect that a resummation of leading $\ln (1/x)$ is necessary to
obtain reliable predictions~\cite{diff}. This resummation 
of terms proportional to $\alpha_s^n\ln^{2n}(1/x)$ for the polarized 
quark and gluon structure functions has been performed in~\cite{ber}. 
The resummation effects are even more pronounced than in the unpolarized 
gluon structure function, which contains only singular terms like
$\alpha_s^n \ln^{n}(1/x)$. For the case of polarized diffraction, one 
expects therefore that the ratio of perturbative to non-perturbative 
contributions is more favourable than in the unpolarized case for two reasons:
the suppression of the relevant non-perturbative three particle cuts and 
the enhancement of the perturbative 
cross section due to large logarithmic terms. 

A different perturbative approach to diffraction in 
spin-dependent deep inelastic scattering has been studied in~\cite{golo}. 
In this approach, it is assumed that the diffractive process is mediated 
by the emission of a non-perturbative pomeron off the proton, the  
pomeron subsequently splits into a $q\bar q$ pair, which interacts with 
the virtual photon. The spin dependence of this reaction is induced by  
a spin-flip component of the pomeron emission vertex and the spin 
dependence of the
pomeron-$q\bar q$ coupling. Using model estimates 
for these spin-dependent couplings, it was found that spin asymmetries 
in diffractive deep inelastic scattering could amount 
up to several percent. 

The spin dependence of diffractive vector meson production can as 
well be predicted in the perturbative
two gluon exchange model. Such studies have been carried out
so far~\cite{vecmes}
 for diffractive photoproduction of $J/\psi$ mesons and 
for diffractive leptoproduction of $\rho$ mesons
at HERA. The observable 
electon-proton spin asymmetries were in both cases found to be only of the 
order of a few per mille, which is below the anticipated statistical 
errors of a measurement even with an integrated luminosity of 1000~pb$^{-1}$.
A study of the spin dependence of the diffractive vector meson 
production cross sections thus appears not to be feasible.

In summary, the spin dependence of 
diffractive deep inelastic scattering and diffractive 
vector meson production have been studied during the workshop. It 
turned out that in particular polarized 
diffractive deep inelastic scattering 
could provide a crucial test for the perturbative description 
of diffraction. It is expected that this observable receives only 
small non-perturbative contributions and is thus largely predictable 
within perturbative QCD. Within the perturbative framework, one 
would moreover anticipate that the polarized diffractive cross section 
at small $x$ displays even stronger logarithmic enhancement than its
unpolarized counterpart. Its measurement would therefore be a crucial test 
of the perturbative interpretation of diffraction in deep inelastic scattering.

\section{Effects at high $Q^2$}
\vspace{1mm}
\noindent
Recently the HERA collaborations~\cite{highq2}
 have reported a significant
excess over Standard Model 
expectation of events produced  in the region of $Q^2$ larger than
10000 GeV$^2$. A statistical fluctuation of the data is not yet excluded
as an explanation, and   
the observed effect  will need  confirmation from larger statistics 
data samples collected in 1997 and thereafter.
 However, if this excess is real, it
could be a first sign of new physics, such as leptoquarks, 
squarks with  R-parity violation or contact interactions.
Alternatively the excess could be a result of our incomplete knowledge on the 
proton structure and/or its perturbative evolution.
On all these topics a plethora of papers has been produced since the 
announcement of the excess.

In this workshop the impact of a polarized HERA  on 
the study of this effect was considered. A general study was made based on the 
contact interaction formalism~\cite{virey}, 
which in principle can mimic any new
physics manifestation in $eq\rightarrow eq$ scattering. It was 
demonstrated that a fully polarized HERA would be very instrumental 
in disentangling the chiral structure of the new interactions. 
Assuming the availability of
both positron and electron beams, and for each beam a data sample 
of 250 pb$^{-1}$ with the spin aligned and anti-aligned to the proton spin,
seven different asymmetries were formed, including two parity 
violating ones, four parity conserving ones, and a mixed asymmetry
(for details see \cite{virey}).
With these data samples the asymmetries are sensitive to
contact interactions to scales
larger than 7 TeV (95\% C.L.),
 and give better limits than equivalent data samples 
with unpolarized beams or with  polarized lepton beams only. 
In the presence of a signal these different
combinations of cross sections into the seven different asymmetries
allow a complete identification of the chiral structure of the new
interactions, i.e.~whether the interactions are LL, RR, LR or LR or 
a combination of those (where L and R denote the left and right handed 
fermion helicities  for the lepton and quark respectively).

Also in  more specific scenarios which imply the production of new particles 
to explain the HERA excess, polarized HERA will play a pivotal role.
E.g.~for the leptoquark production 
scenario, the compatibility of the HERA result
with results from pion decays and from $(g-2)_{\mu}$ experiments,
 induces large
effects in the parity violating asymmetries. A special case 
was studied in~\cite{ellis} for stop squark production off 
strange and down quarks in the proton within an R-parity violating 
supersymmetric scenario.
It was shown that 
one can take advantage of our knowledge of the polarized quark distributions
in the proton, which in this case are different for down and strange quarks, 
to differentiate between different possible scenarios
from the measured production rates at a polarized HERA.

At this time it is not established that the reported excess invokes new 
physics, and more conservative scenarios, particularly 
those concerned with the structure of the proton, have been explored as well. 
The present Standard Model cross section estimate is based on 
parton densities measured at low $Q^2$ and high 
$x$ which are evolved to the high 
$Q^2$ region. Within the framework and limits of present day modern 
global fits, the expected uncertainty on the parton distributions is 
claimed to be of the order of 10\%. However, it has been pointed out 
that the low $Q^2$ data do not exclude the presence of additional 
components in the high-$x$ range, such as additional gluons, charmed 
quarks or meson-cloud effects. A particularly interesting 
possibility is  the effect induced by  QCD 
instantons~\cite{kochelev}. 
 Non-perturbative instanton fluctuations
describe the quantum tunneling between different gauge rotated classical
vacua in QCD (see e.g.~\cite{instanton}).
Due to the quark helicity flip at the quark-instanton vertex, the 
contribution to the spin-dependent cross sections of instantons is 
very different from the one of the perturbative quark-gluon vertex.
Furthermore, in the instanton liquid model~\cite{instanton} 
the  contribution of instantons to the  proton structure
is expected to  become increasingly more important with increasing
$Q^2$~\cite{kochelev}.  
The instanton liquid model yields definite predictions~\cite{kochelev}
for both 
unpolarized and polarized structure functions at large $Q^2$, and the 
prediction in the unpolarized case is consistent with the 
presently observed excess.
A measurement of the 
spin-dependent cross sections at large $Q^2$ with at a polarized 
HERA would immediately and 
unambiguously test this instanton interpretation.

In all, if  HERA continues to  produce more events at high $Q^2$ than 
expected, a polarized HERA will be essential for our 
complete  understanding  on the origin of this effect.

\section{Conclusions}
The operation of HERA with polarized beams appears as a natural continuation 
of the successful physics programme of HERA, both in the unpolarized
sector with H1 and ZEUS, as well as in the polarized sector with HERMES.
It will  allow to make unique 
measurements in polarized  deep inelastic scattering as well as
photoproduction at centre of mass energies of a few hundred GeV. 
HERA is the only machine in the world where this could be realized, and
a rich programme of spin-dependent physics will emerge if data samples
corresponding to a few hundred inverse picobarns can be collected.
The high energy polarized proton beam also leads to the 
opportunity of a fixed target polarized $pp$ experiment to study single 
and double spin asymmetries at a CMS
energy of about 40 GeV~\cite{nowak}.

Several potential measurements have been studied in great detail 
in this workshop, and many new channels have been tackled. The necessity 
for low-$x$ measurements of the structure functions, and 
determination of the polarized gluon
distribution $\Delta G$ have 
been wildly advocated by the spin physics community over the last 2 years.
HERA can play a pivotal role in this field since it is able 
 to give conclusive insight on both of these issues. 
No other accelerator is able to provide data for the measurement of the 
polarized structure functions in the region covered by HERA.

HERA will also contribute to the flavour decomposition of the quark spin
distributions, spin transfer in quark fragmentation, spin effects in 
diffractive scattering, and the very intruiging possibility
to measure polarized parton distributions in the photon. 
Finally, a polarized HERA will be very instrumental in the study and
interpretation of possible deviations from the Standard
Model expectation in the  high-$Q^2$
region.
The physics scope can be considerably extended if polarized $^3$He beams
would become available as well, and if the present detectors could be
further instrumented in the proton remnant region.

The results obtained during both workshops in 1996 and 1997 clearly
show
 that a polarized proton beam at HERA, in conjunction with the
already successfully operating polarized electron beam, would
undoubtedly pave the way to significantly improve our present
 understanding  of the spin
structure in hadrons, which is one of most interesting challenges of
QCD these days. Polarized beams at HERA constitute a very 
strong physics case that should be considered 
for the future planning of research at DESY.

\section*{Acknowledgements}
\vspace{1mm}
\noindent
It is a pleasure to thank all participants
 of the ``Physics with Polarized 
Protons at HERA''-workshop for the 
contributions during the 
meetings and to the proceedings.

\end{document}